\documentclass[pra,aps,twocolumn,eqsecnum,showpacs]{revtex4}

\usepackage{dcolumn}
\usepackage{amsmath}
\usepackage{graphics}

\begin{document}
\title{Multiple light scattering on the $F = 1 \rightarrow F'  = 0$ transition\\ in a cold and high density $^{87}$Rb vapor}

\author{S. Balik, A.L. Win and M.D. Havey}
\affiliation{Department of Physics, Old Dominion University,
Norfolk, VA 23529} \email{mhavey@odu.edu}

\author{A.S. Sheremet, I.M. Sokolov and D.V. Kupriyanov}%
\affiliation{Department of Theoretical Physics, State
Polytechnic University, 195251, St.-Petersburg, Russia}

\date{\today }

\begin{abstract}
We report an experimental study of near resonance light scattering
on the $F = 1 \rightarrow F'  = 0$ component of the $D_2$ line in
atomic $^{87}Rb$.  Experiments are performed on spatially
bi-Gaussian ultracold gas samples having peak densities ranging
from about $5 \cdot 10^{12} - 5 \cdot 10^{13}$ atoms/cm$^{3}$ and
for a range of resonance saturation parameters and detunings from
atomic resonance.  Time resolution of the scattered light
intensity reveals dynamics of multiple light scattering, optical
pumping, and saturation effects. The experimental results in steady-state are compared qualitatively with theoretical models of the light scattering process.   The steady-state line shape of the excitation spectrum is in good qualitative agreement with these models.
\end{abstract}


\maketitle%

\section{Introduction}
The interaction of light with dense atomic gases is a vigorous
area of research in quantum optics
\cite{HaveyCP,LabeyrieReview,LPLReview,CommentAMO,KaiserHavey,Metcalf,Grimm,Pethick,Giorgini,Bouwmeester,Lukin,Milonni,
Marangos,Hau,Braje,Jin,Campbell,Ye,Rolston,Killian,Carr,Weidemuller,Lukin1,Kuzmich}.
Part of this interest stems from the interdisciplinary nature of the field, and with the large
number of fundamentally important results and potential
applications that have emerged.  Recent efforts have ranged from
basic studies of light localization in disordered systems
\cite{Anderson,Akkermans1,KupriyanovLightStrong,Fofanov,JETP,Wiersma1,Chabanov1,Maret1,Maret},
including atomic gases, to investigation of cooperative scattering \cite{Scully,Svidzinsky,KaiserCoop,KaiserCoop2,KaiserCoop3,KaiserCoop4,KaiserCoop5,KaiserCoop6}.  Areas of experimental
and theoretical research with both fundamental motivations and
possible applications include searches for atomic physics based
random lasing
\cite{Cao,Wiersma,Conti,KaiserRandom1,KaiserRandom2,KaiserRandom3} and quantum memories for quantum information and
communications \cite{Nielsen,Briegel,Duan}. Single photon optical memories in dense atomic
gases may potentially be derived from a number of approaches,
including development of subradiant atomic-photonic modes \cite{KaiserSubradiant}, and
extensions of electromagnetically-induced-transparency (EIT) based
approaches to novel two-photon optical schemes at higher
densities \cite{Lukin,Marangos,Hau,Lukin1,Kuzmich,Datsyuk1,Datsyuk2}.

We have ongoing experimental and theoretical research efforts
focused in part on developing quantum memories using either
two-photon EIT based approaches on one hand \cite{Datsyuk1,Datsyuk2,qm1,qm2,qm3}, and on the possible
formation of subradiant single photon modes on the other \cite{KupriyanovLightStrong,Fofanov,JETP}. To
obtain formation of subradiant modes is a challenging experimental
enterprise, and requires atomic densities $\sim 10^{14}$
atoms/cm$^{3}$, in order to achieve high orders of multiple light
scattering in the sample.  In addition, dynamical processes which
either dephase the multiply scattered light or which lead to
reduction in the scattering cross-section should be well
understood.  The current program is focused on nearly optically
closed hyperfine transitions associated with the $D_2$ transition
in ultracold $^{87}Rb$. These two transitions are the $F = 2
\rightarrow F' = 3$ main optical trapping component and the $F = 1
\rightarrow F' = 0$ transition arising from the lower energy
ground state hyperfine component.  We have reported elsewhere our
experimental research associated with light scattering on the $F =
2 \rightarrow F' = 3$ transition \cite{Balik23}.

In the present paper we report experimental and theoretical investigation of the
$F = 1 \rightarrow F' = 0$ transition in $^{87}Rb$.  Overall, our studies include
examination of the roles of atomic density, optical saturation,
and detuning of probe radiation from optical resonance on the $F =
1 \rightarrow F' = 0$ transition.  We will see that, for this transition, the
debilitating effects of Zeeman optical pumping play an essential
role in all aspects of the experiments.  In the present paper, we concentrate primarily on steady state optical excitation and how the light scattering signals depend in that case on probe detuning and atomic density.  In the following sections
we first provide some details of our experimental approach.  This
is followed by presentation of our experimental results and
comparative discussion on the basis of theoretical models of the light scattering process.

\section{Experimental configuration}

\begin{figure}[htpb]
\includegraphics{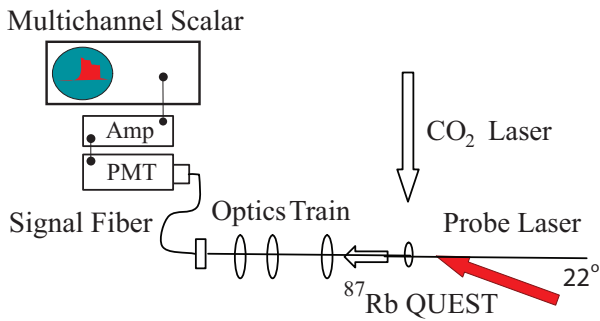}%
\caption{Schematic diagram of the experimental apparatus.  In the
figure, PMT refers to an infrared sensitive photomultiplier tube
and Amp refers to a fast preamplifier. MOT stands for magneto optical trap,
while QUEST is an abbreviation for quasi electrostatic trap.}
\label{fig1}%
\end{figure}

\begin{figure}[htpb]
\includegraphics{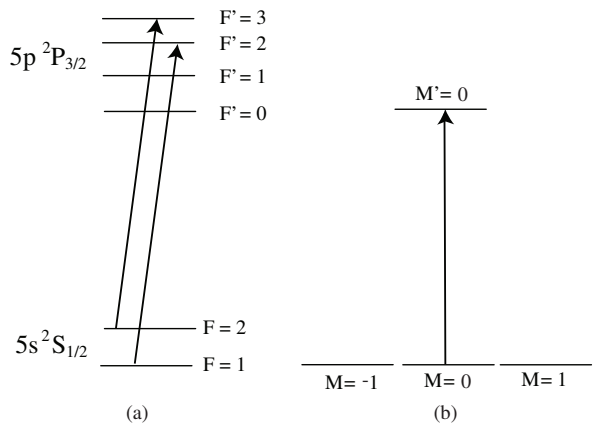}%
\caption{Energy level diagrams illustrating the experimental
scheme.  (a) The MOT and repumping transitions used in the
present experiment.  (b) Probe excitation with linearly polarized
light (z-direction) used in the present experiment.  In part (b)
of the figure, the Zeeman levels have been broken out to show the
role of Zeeman optical pumping out of the F = 1, M = 0 state to
the F = 1, M = $\pm$ 1 states.}
\label{fig2}%
\end{figure}

A schematic diagram of the experimental apparatus used in the
measurements is shown in Fig. 1, while the particular optical
transitions of interest associated with the atomic $^{87}Rb$ $D_2$ line are
shown in Fig. 2. As illustrated in Fig. 1, the central part of the
experimental apparatus is a magneto optical trap (MOT) which serves to form and confine
cold $^{87}Rb$ atom samples. The MOT is a standard vapor-loaded
trap formed in a vacuum chamber with a base pressure $\sim
10^{-9}$ Torr. The six MOT beams are derived from a single
external cavity diode laser (ECDL) with the grating arranged in a
Littrow configuration. The main master diode laser is frequency
locked to a saturation absorption feature produced in a room
temperature Rb vapor cell. The master laser power is increased by
injecting the master output into slave laser.  The arrangement
provides more than 20 mW of trapping light in laser beams of cross
sectional area $\sim$ 2 $cm^{2}$ . The slave laser output is
switched and spectrally shifted as required with an acousto
optical modulator (AOM) to a frequency set at about 18 MHz below
the $^{87}Rb$ $F=2\to F'=3$ trapping transition.  The repumper
laser is also an ECDL of the same basic design as the main MOT
laser, and is locked to the $F=1\to F'=2$ hyperfine transition.
The repumper delivers a beam of maximum intensity $\sim$ 0.6
$mW/cm^{2}$ and is delivered along the same optical path as the
main trapping laser beams. Switching of the repumper laser is also
controlled with an AOM.

In the experiments reported here, the cold atom sample is
initially produced in the higher energy F = 2 level. Direct
absorption imaging measurements of the peak optical depth on the
$F=2\to F'=3$ transition yielded, for this sample, $b_{o}$ $\sim$
10 in a Gaussian radius of $r_{o}$ $\sim$ 0.45 mm.
However, the main sample production goal is to transfer a
significant number of the trapped atoms to a carbon-dioxide laser
($CO_{2}$) based far off resonance optical dipole trap.  The 100 W
$CO_{2}$ laser based trap operates at a wavelength of 10.6 $\mu m$
and is deeply in the quasistatic trapping regime.  This laser is
focussed to a radial spot size of $\sim$ 55 $\mu m$, and a
corresponding Rayleigh range of $z_{R}$ $\sim$ 750 $\mu m$.  The
$CO_{2}$ laser focal zone is overlapped with the MOT trapping
region, while application of the laser beam itself is controlled by
a 40 MHz AOM. The atom sample formed in the MOT is compressed and
loaded into the quasistatic dipole trap (QUEST) by detuning the
MOT master laser 60 MHz to the low frequency side of the trapping
transition, while simultaneously lowering the repumper intensity
over an order of magnitude.  The resulting temporal dark spot MOT
loads the atoms predominantly into the lower energy F = 1
hyperfine component. The result of this procedure is transfer of
about 15 $\%$ of the MOT atoms to the QUEST.  It is important to
note that this transfer efficiency is determined after a QUEST
holding period of about 1 second, during which the atomic sample
naturally evolves towards thermal equilibrium (this happens
through elastic collisions between the confined Rb atoms).
Auxiliary measurements of the QUEST principal characteristics
after the 1 s hold period, by absorption imaging, parametric
resonance \cite{parametric1}, and the measured number of atoms transferred show a
sample with peak density about 5 $\cdot$  10$^{13}$ $atoms/cm^{3}$
and a temperature of$\sim$ 65 $\mu K$.  The 1/e lifetime of the
confined atoms is longer than 5 s, and is limited by background
gas collisions. The residual magnetic field in the sample area,
when the MOT quadrupole field is switched off, is estimated to be
less than a few mG.

In the main experimental protocol, a probe beam tuned in the
spectral vicinity of the $F = 1 \rightarrow F' = 0$ nearly closed
transition is directed towards the sample, and the resulting
scattered light signals collected as illustrated schematically in Fig. 1.
The probe laser is of the same design as the repumper laser, has a
bandwidth $\sim$ 3 MHz,  and is switched and directed by an acousto
optical modulator towards the sample. Because of constraints on the
vacuum chamber geometry, the linearly polarized probe beam is directed
(see Fig. 1) at an angle of approximately 30 degrees away from the
fluorescence collection direction.  The probe beam is also directed
downwards towards the sample at an angle of 22.5 degrees (as shown in Fig. 1). Finally, the
sample fluorescence is collected without regard to light polarization;
as light scattering is dominated by the $F = 1 \rightarrow F' = 0$
transition we expect the scattered light to be mainly unpolarized
(very small contributions from the quite far off resonance $F = 1 \rightarrow F' = 1,2$
transitions are in general polarized in the single scattering limit).

To close this section, we point out that in some of the
experiments reported here the atomic density was varied over a
factor of about 10.  This was accomplished by allowing for a period of
ballistic expansion of the cloud after the QUEST was turned off.
The atomic sample temperature is known (by ballistic expansion measurements), so this procedure allows
the peak density or the peak optical depth to be determined.   As
the sample is well approximated by a two-axis Gaussian atom
distribution \cite{QUESTFormulas}, the two Gaussian radii and the
peak atom density (or the total number of atoms in the sample),
are sufficient to determine the two peak optical depths
characterizing the sample. We summarize in Table 1 the peak
transverse optical depth $b_{t}$ the peak atom density at the
center of the sample $n_{o}$, the transverse Gaussian radius
$r_{o}$, and the longitudinal Gaussian radius $z_{o}$.  The
optical depth refers here to that of the nearly closed $F = 1
\rightarrow F' = 0$ hyperfine transition, which has a total resonance
light scattering cross section of $3.23 \times 10^{-10}$ $cm^{2}$.

\begin{table}
\begin{tabular}{|c|c|c|c|}
  \hline
  Peak $b_{t}$ & $n_{o}$ (atoms/cm$^{3})$ & $r_{o}$ ($\mu m$)& $z_{o}$ ($\mu m)$\\
  \hline
  40 & 5.0 $\times 10^{13}$& 9.8 & 248 \\
  28 & 2.5 $\times 10^{13}$ & 13.8 & 248 \\
  20 & 1.2 $\times 10^{13}$& 19.5 &  248 \\
  13 & 5.1 $\times 10^{12}$& 30.4 & 249 \\
  \hline

\end{tabular}
\caption{QUEST parameters relating the peak transverse optical
  depth on the $F = 1 \rightarrow F' =0 $ transition
  to the maximum sample density and the Gaussian radii of the
  atomic cloud.}
\end{table}

\section{Results and discussion}
In this section we present results associated with on-resonance
light scattering, where on-resonance refers to the bare-atom $F =
1 \rightarrow F'  = 0$ hyperfine resonance frequency, and with
variations around that frequency.  We emphasize that the
measurements are made at a fixed single angle with respect to the
incident probe laser; the geometrical setup is described in the
previous section.  With this in mind, we first expect that the
scattered light will be unpolarized, as the excited level has $F'$
= 0.  However, because of the high optical depth of the sample,
the scattered light intensity should show important angular
dependence \cite{JETP,qm1,Scaling}. For instance,
light scattered in the near forward direction shows a minimum in
the spectral variations near the resonance line center. This can
effect the time evolution of the scattered light signals as well
\cite{qm1}. On the other hand, light scattered in the
backwards direction shows an enhancement due to the coherent
backscattering effect.  This spatially cone shaped feature depends
critically on probe laser intensities near saturation and also on
the spectral detuning from atomic resonance.

\subsection{On-resonance scattering}

\begin{figure}[htpb]
\includegraphics{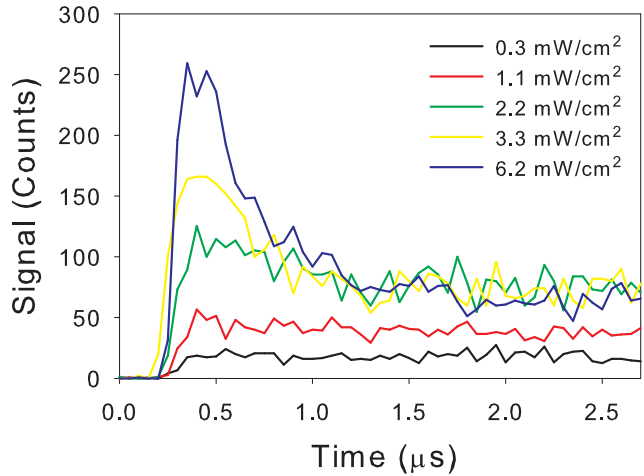}%
\caption{Time evolution of the probe scattered light intensity on
the $F = 1 \rightarrow F'  = 0$ hyperfine transition.  The atomic
density is at its peak level for these measurements.  Signals are
shown for several different probe laser intensities. The role of
optical pumping is apparent at the larger probe intensities. Probe
detuning from resonance $\Delta$ = 0.}
\label{fig3}%
\end{figure}

\begin{figure}[htpb]
\includegraphics{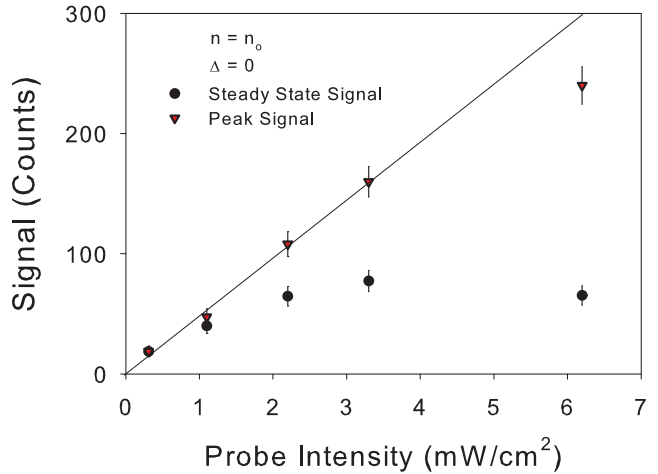}%
\caption{Variation of the peak and steady state signal levels as a
function of the probe laser intensity. Atomic density is at is
maximum value $n_o$ for these data. Probe detuning from resonance
$\Delta$ = 0.}
\label{fig4}%
\end{figure}

We start by presenting in Fig. 3 the results of typical
measurements of the time dependence of light scattered from the
atomic sample.  These measurements are made for a maximum density
sample, which in this case has a transverse optical depth of $b_t$
$\sim$ 40, and a peak atom density at the center of the sample of
5.0 $\cdot$ 10$^{13}$ atoms/cm$^{3}$. These measurements are made
for a range of probe laser intensities, which for the most part
are well below the saturation intensity $I_{sat}$ $\sim$ 14 $mW/cm^{2}$
for linearly polarized excitation of the $F= 1 \rightarrow F'  = 0$ transition.  We see for higher probe
laser intensities a quite rapid buildup to a peak intensity
followed by a decay of about 1 $\mu$s to a steady state.  The
build up time is less than about 100 ns, and is limited by the
turn-on time of the switching AOM used to control application of
the probe laser to the sample.  We also note that as the probe
laser intensity is reduced, the transient peak is reduced, and for
the lowest probe laser intensities the scattering signal smoothly
rises to a steady value.  The overall behavior is summarized in
Fig. 4, where the nearly linear growth of the peak in the
scattering signal, and the leveling off of the steady state
signal is shown.   This general behavior of the transient response and the variations of the
response with probe laser intensity was found to be qualitatively the same for
a range of densities from 0.1$n_o$ up to the peak density $n_o$ (see Table 1).

\begin{figure}[htpb]
\includegraphics{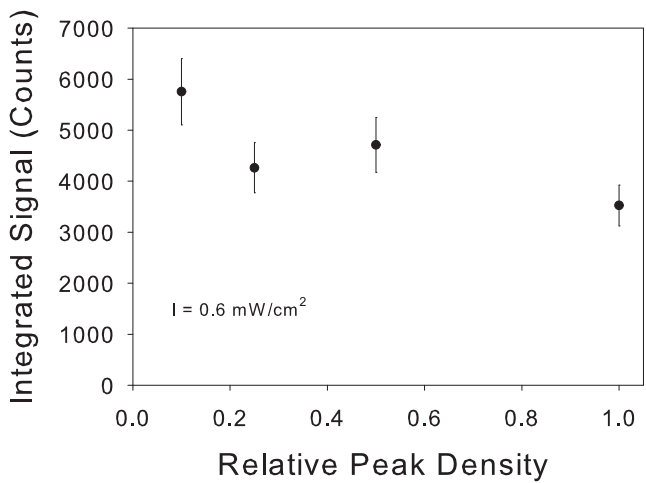}%
\caption{Variation of the total scattered signal intensity due to
systematic changes in the atomic density. Probe detuning from
resonance $\Delta$ = 0.}
\label{fig5}%
\end{figure}

\noindent As the maximum laser intensity used in these experiments is well below the optical saturation intensity, we can understand the transient behavior to result from Zeeman optical pumping in the ground level due to elastic Raman transitions to the $F = 1, M = \pm 1$ states. This optical pumping does not completely deplete the $F=1, M = 0$ state because multiple elastic light scattering allows for a build up of radiation within the atomic sample. This gives rise to optical pumping by the diffuse light in the sample, which competes with direct Zeeman optical pumping by the probe beam and results in a nonzero steady state level for the scattered light signal as seen in Fig. 3 and Fig. 4.  This process as considered for a wide range of densities and laser intensities will be treated in a later report.  In the present paper we are concerned with the steady state signals for the case where the probe intensity is so weak, and the number of scattered photons so few, that there is no mesoscopic rearrangement of populations in the F = 1 level during realization of a single sample.   The main results of this paper, as presented in this and the following section, are recorded under those conditions.

We present in Fig. 5 measurements of the variation of the total scattered light intensity from the $F = 1 \rightarrow F'  = 0$ hyperfine transition as a function of systematic changes in the atomic density. These measurements were made at very low probe laser intensity, under conditions selected so that there was no perceptible optical pumping, as in the lowest intensity results shown in Fig. 3 and Fig. 4.  In Fig. 5 we see that as the density decreases, the overall intensity of the scattered light increases.   A similar behavior \cite{Balik23,Scaling} has been observed for measurements on the $F = 2 \rightarrow F'  = 3$ hyperfine transition of $^{87}Rb$.   The effect is due to the collective nature of near resonance light scattering from a high density and cold atomic gas.  Because the optical depth is so large for the highest density, light scattering occurs primarily from the outer surface of the sample.  Relatively few atoms then contribute to the sample.  For lower density the light can penetrate more deeply and a relatively larger number of atoms participate, leading to a larger signal.

\subsection{Variations with spectral detuning from resonance}
We now consider the variations of the scattered light signals as a function of detuning of the probe laser frequency from bare atomic resonance, where $\Delta = 0$.   These data are recorded under conditions of a very weak probe laser, so that Zeeman optical pumping is negligible. For the steady state regime (as in the lowest probe intensity results of Fig. 3), this dependence is shown in Fig. 6.   There we see that the spectral variations show a clear resonance behavior in a range of 3 to 4 $\gamma$ about $\Delta = 0$, where $\gamma$ is the $\sim$ 6 MHz natural width of the atomic resonance.  The solid line is a Lorentzian line profile, which well describes the resonance line shape, and yields a full width at half maximum of slightly larger than 10 MHz.  A full-width greater than the natural width is expected, due to absorption broadening and due to the dipole dipole interaction of the atoms under high density conditions. Selection of the Lorentzian line shape is arbitrary, but gives a decent fit to the data and a consistent way of estimating the full width under different experimental conditions.  Within the spread of the data, a fit using a Gaussian line profile yields essentially the same full width at half maximum.

\begin{figure}[htpb]
\includegraphics{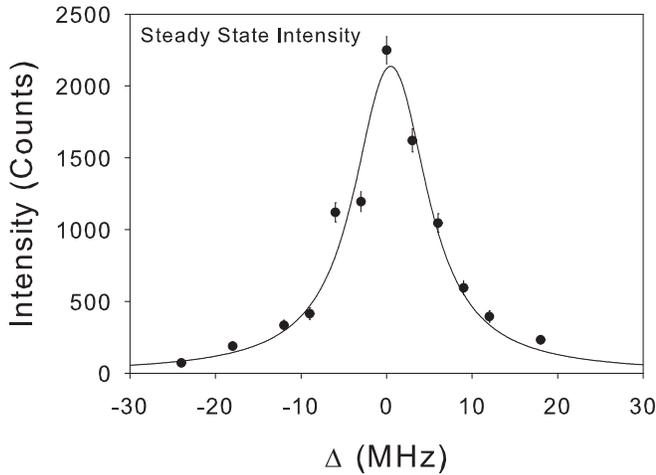}%
\caption{Near-resonance spectral variation of the total scattered light intensity in the
spectral vicinity of the $F = 1 \rightarrow F'  = 0$ hyperfine
transition. }
\label{fig6}%
\end{figure}

To further analyze the experimental data, we have extracted the full width at half maximum of the spectral profile as a function of time \cite{Balik23}.   With reference to the lowest intensity results of Fig. 3, this is equivalent to making a slice on the time axis of 100 ns and recording the spectral profile as a function of detuning in this time window.  Repeating this procedure for the full sequence of data over the 2.5 $\mu s$ range of the probe excitation and decay signals yields the excitation spectrum shown in Fig. 7.   There we see that the line width is very large for short times, rapidly decays to its steady state level (as in Fig. 6), and then sharply decreases again after the probe pulse is extinguished.  For the short-time turn on of the excitation pulse single scattering dominates, and the line width can be estimated to be $\Delta_o$ = $\gamma /2$$\sqrt{b_o - 1}$, where $b_o$ is the on resonance transverse optical depth through the center of the cloud.  For the experiments reported here $b_o$ = 40, giving a full width of $\Delta_o$ = $\gamma$$\sqrt{b_o - 1}$ $\sim$ 37 MHz, in very good agreement with the short time value of Fig. 6.

Upon turn off of the probe pulse the spectral width of the excitation spectrum decreases from its steady state value of about 10 MHz to a value on the order of the natural width.   As in the spectral response upon turn-on of the prober pulse, the line shape here is also well fit by a Lorentzian form. However, we point out that the spectral width of the probe laser itself is about 1 MHz, and has a measured Gaussian power spectrum.  This means that the widths determined by these measurements are slightly larger than that determined by the physical processes involved.  In fact, the longest lived mode \cite{Balik23} for these samples corresponds to the so called Holstein mode, or the longest lived diffusive mode for the sample under study; the lifetime of this mode can be significantly smaller than the natural width of the transition.

\begin{figure}[htpb]
\includegraphics{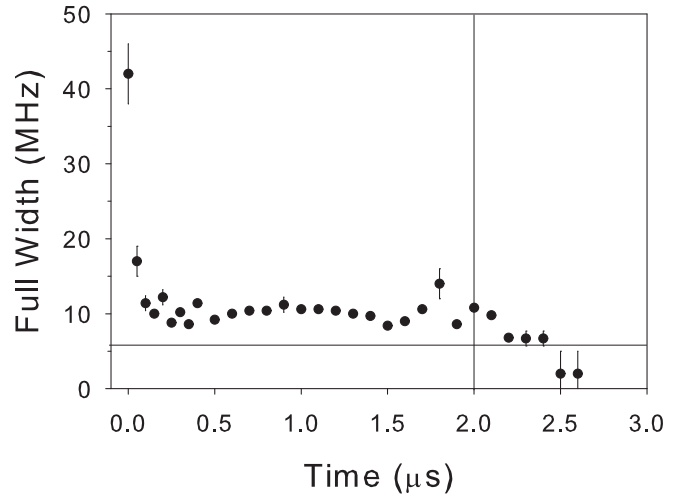}%
\caption{Time evolution of the probe scattered light intensity on
the $F = 1 \rightarrow F'  = 0$ hyperfine transition.  The atomic
density is at its peak level for these measurements.}
\label{fig7}%
\end{figure}

\section{Theory}
In this section we present theoretical calculations for comparison with the steady state line shape of the excitation spectrum. In general, microscopic calculations for the cooperative scattering process for a macroscopic collection of atoms with a degenerate ground state is rather difficult because of the rapidly rising number of equations to be solved $d_eN d_{g}^{N-1}$, where $d_e$ is the degeneracy of the atomic excited state and $d_g$ is the degeneracy of the ground state. Here N is the number of atoms considered in the calculations.  This makes practically impossible exact microscopic calculations for the scattering process and for its time-dependent fluorescence dynamics. In this section we present the results of our calculations of the scattering spectrum in the steady state regime, which are based on two complementary models of the self-consistent and approximate microscopic approaches, see Ref. \cite{KupriyanovLightStrong}, and discuss the results of our numerical simulations in the context of the data presented in the experimental part of the paper.

\begin{figure}[tpb]
\includegraphics{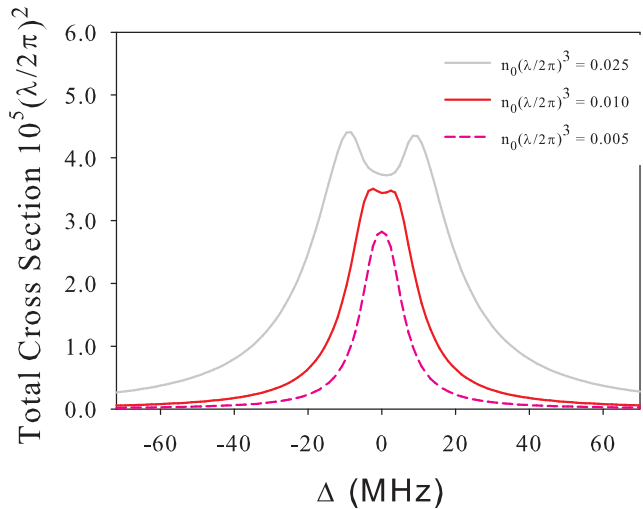}%
\caption{The scattering cross section calculated in the Debye-Mie model for an atomic sample with radius $30\; \mu m$. Atoms fill a sphere homogenously with the density varying from $0.005$ to $0.025$. }
\label{fig8}%
\end{figure}

The self-consistent approach accepts the description of the scattering problem on the level of the macroscopic (i.e. mesoscopically averaged) Maxwell theory, where the dielectric susceptibility obeys an equation resulting from the self-consistent dynamics of the driving (probe) field and of the atomic dipoles. The crucial assumption is that groups of closely located atomic dipoles respond to the field cooperatively such that the longitudinal dipole-dipole interaction can be incorporated into the Lorentz-Lorenz local-field correction. For a low intensity of the driving field (where optical pumping can be neglected) the atoms equally populate the Zeeman sublevels and the medium is isotropic and can be parameterized by a single dielectric constant. If the atoms are homogeneously distributed in a spherical volume, then it is possible to express the scattering cross section by the standard solution of the Debye-Mie problem. In Fig. \ref{fig8} we reproduce the spectral dependencies of the total cross section calculated for such a spherically symmetrical atomic system. The chosen radius of the sphere $30\; \mu m$ is roughly scaled by the spot size where the probe beam crosses the atomic sample. We use the densities essentially less than in the peak point of the cloud to emphasize the importance of the contribution from the tail area of the Gaussian distribution in the experimental observation of the scattering process. The plotted dependencies demonstrate the smoothed spectral profile with a bandwidth qualitatively in agreement with the experimental data of Fig. \ref{fig6}. Note that there is a slight asymmetry in the spectral behavior related with the density effects which is a consequence of the spectral asymmetry of the dielectric permittivity. Because of the essential differences in geometries of this model and the experiment, which is a prolate double Gaussian distribution in experiment and spherically homogeneous in theory, we cannot make direct quantitative comparison of experimental and theoretical results but we can point out at least the qualitative agreement between both the data sets.  We finally point out that the near resonance theoretical behavior is structured as a competition of longitudinal red shift and cooperative radiation blue shift but this effect is very likely masked by the uncertainty in experimental data.

We note that the time dynamics associated with the fluorescence decay can be also modeled in the framework of the self-consistent calculation scheme. However, this would require the complete Monte-Carlo simulations of the process with the Green's function formalism, see Ref. \cite{LPLReview}. The scattered energy of the light scattered from a dense atomic cloud in the steady state regime emerges mainly from its surface area and the transient deviations for the widths of the fluorescence spectra reproduced in Fig. \ref{fig7} could be verified by such a Monte-Carlo simulated and spectrally sensitive self-consistent calculations.

As we commented above, the microscopic calculations can be only approximately done for the considered transition. Here we reproduce the results of our calculations based on the general formalism of the quantum scattering theory previously developed in Ref. \cite{KupriyanovLightStrong} and present the data for the case of a sample consisting of one hundred atoms. For the typical experimentally attained atomic densities, which are less than $n_0\lambdabar^3\lesssim 0.1$ in its peak value, in order to resolve the quasi-energy structure, associated with the longitudinal dipole-dipole interaction, it would be enough to take into consideration only a few nearest neighbors in the vicinity of each selected scatterer. This scheme were tested by us on a small number of atoms and it demonstrated reliable convergence, which reduces the number of equations to be solved to  $d_eN d_{g}^{n-1}$, where $n$ is the number of the neighboring atoms retained in the calculation.

\begin{figure}[tpb]
\includegraphics{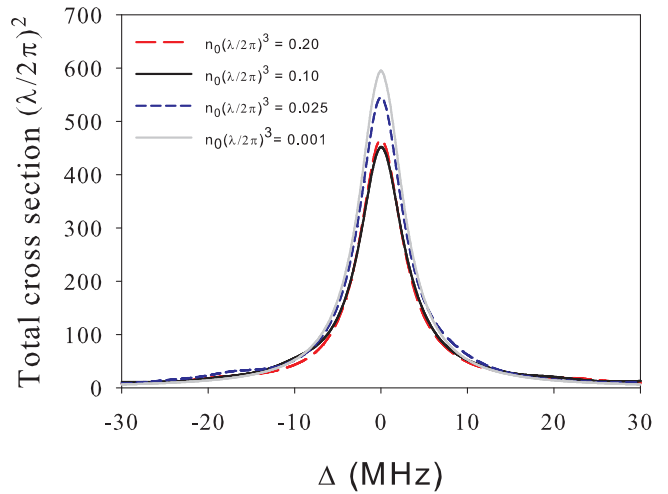}%
\caption{The scattering cross section calculated microscopically for an ensemble consisting of one hundred atoms. Atoms have a spherically symmetrical Gaussian distribution with the peak density varying from $n_0\lambdabar^3=0.001$ to $0.2$ and populating only one Zeeman sublevel. }
\label{fig9}%
\end{figure}

In Fig. \ref{fig9} we show the results of our calculations, which were done for $n=4$ and for all the atoms populating one Zeeman sublevel. Let us point out here again that because of high degeneracy of the ground state our results cannot be statistically averaged over all the populated initial states. That means we cannot directly compare them on a one-to-one basis  either with experiment or with the self-consistently calculated data. Instead, we rely on qualitative comparison among the results.

First, as was explained in Ref. \cite{KupriyanovLightStrong}, because of the presence of elastic Raman scattering channels there is no visible signature of either super- or sub-radiant Dicke-type exciton modes for the $F=1\to F'=0$ transition. In contrast with $F=0\to F'=1$ transition (where such states are normally predicted and discussed) in the $F=1\to F'=0$ case all the resolvent poles are described as the resonances with a typical line width of about $\gamma$.  Thus the presence of the  cooperative longitudinal and radiative interactions manifest themselves in the smoothed variation of the scattering spectrum associated with disorder and with the configuration dependence. The slight modulation of the experimental spectrum, which is suggested in the experimental data shown in Fig. \ref{fig6} can be not only experimental uncertainty but can be also manifestation of the cooperativity and disorder effects. Second, we can point out that because of competition between near and far field interactions the microscopically calculated spectra have an asymmetry in respect to the atomic resonance line. The latter observation is in qualitative agreement with the predictions of the self-consistent model, which were pointed out above.

\section{Conclusions}
We have presented detailed experimental and theoretical results associated with
scattering of light from an ultracold and highly dense
gas of $^{87}Rb$ atoms.  For radiation tuned in the spectral
vicinity of the $F = 1 \rightarrow F'  = 0$\textbf{} hyperfine component of
the $D_2$ line, we have studied the atomic density, probe laser detuning and probe laser
intensity dependence of the scattered light intensity. The
measured time dependence and steady state responses indicate that
the light dynamics is strongly effected by Zeeman optical pumping in the
lower energy F = 1 hyperfine component.  However, for very low probe laser
intensities, Zeeman optical pumping can be made negligible; this has allowed for study of the steady state regime in the absence of optical pumping.  Good results are obtained in comparison between the experimental results and theoretical calculations.  The strong role of elastic Raman transitions also strongly suggests that this transition is not particularly suitable for searches for Anderson light localization in three dimensions.

\section{Acknowledgments}
We acknowledge Daniel Havey for informative discussions and contributions to the experimental line shape fitting. We appreciate financial support by the National Science Foundation (Grant Nos. NSF-PHY-0654226 and NSF-PHY-1068159), the Russian Foundation for Basic Research (Grant No. RFBR-CNRS 12-02-91056). D.V.K. would like to acknowledge support from the External Fellowship Program of the Russian Quantum Center (Ref. Number 86). We also acknowledge the generous support of the Federal Program for Scientific and Scientific-Pedagogical Personnel of Innovative Russia for 2009-2013 (Contract No. 14.B37.21.1938). A.S.S. would like to acknowledge support from the charitable foundation "Dynasty".


\begin{thebibliography}{99}

\bibitem{HaveyCP} M.D. Havey, Contemp.Phys. 50, 587 (2009).

\bibitem{Metcalf} H.J. Metcalf and P. van der Straten, \textit{Laser
Cooling and Trapping}, Springer, New York, 1999.

\bibitem{Grimm} R. Grimm, M. Weidemuller, and Y. Ovchinnokov,
Adv. Atom., Mol., and Opt. Phys. 42, 95 (2000).

\bibitem{Pethick} C.J. Pethick and H. Smith, \emph{Bose-Einstein Condensation in
Dilute Gases}, Cambridge University Press, Cambridge, UK, 2002.

\bibitem{Giorgini} S. Giorgini, L.P. Pitaevskii, and S. Stringari,
Rev. Mod. Phys. 80, 1215 (2008).

\bibitem{Bouwmeester} Dirk Bouwmeester, Artur Ekert, and Anton
Zeilinger, \emph{The Physics of Quantum Information},
Springer-Verlag, Berlin, Germany, 2001.

\bibitem{Lukin} M.D. Lukin, Rev. Mod. Phys. 75, 457 (2003).

\bibitem{Milonni} P.W. Milonni, \emph{Fast Light, Slow Light, and
Left-handed Light}, Taylor and Francis, New York, 2005.

\bibitem{Marangos} M. Fleishhauer, A. Imamoglu, and J.P. Marangos,
Rev. Mod. Phys. 77, 633 (2005).

\bibitem{Hau} Lene Vestergaard Hau, Nature Photonics 2, 451
(2008).

\bibitem{Braje} D.A. Braje, V. Balic, G.Y. Yin, and S.E. Harris,
Phys. Rev. A 68, 041801 (2003).

\bibitem{Jin} S. Ospelkaus, A. Pe'er, K.-K. Ni, J. J. Zirbel,
B. Neyenhuis, S. Kotochigova, P. S. Julienne, J. Ye, and D. S.
Jin, Nature Phys. 4, 622 (2008).

\bibitem{Campbell} G. K. Campbell, A. D. Ludlow, S. Blatt,
J. W. Thomsen, M. J. Martin, M. H. de Miranda, T. Zelevinsky, M.
M. Boyd, J. Ye, S. A. Diddams, T. P. Heavner, T. E. Parker, and S.
R. Jefferts, Metrologia 45, 539 (2008).

\bibitem{Ye} Jun Ye, S. Blatt, M. M. Boyd, S. M. Foreman,
E. R. Hudson, Tetsuya Ido, B. Lev, A. D. Ludlow, B. C. Sawyer, B.
Stuhl, T. Zelinsky, Int. J. Modern Phys. D 16, 2481 (2007).

\bibitem{Rolston} Steven Rolston, Physics 1, 2 (2008).

\bibitem{Killian} Thomas C. Killian, Science 316, 705 (2007).

\bibitem{Carr} The entire issue, New J. Phys. 11 (2009), focuses
on recent advances and opportunities in ultracold molecular
physics.  See particularly, Lincoln D. Carr and Jun Ye, New J.
Phys. 11, 055009 (2009).

\bibitem{Weidemuller} Matthias Weidem\"{u}ller and Claus
Zimmermann, \emph{Interactions in Ultracold Gases}, Wiley-VCH,
Germany, 2003.

\bibitem{Lukin1} M. Fleischhauer and M. D. Lukin, Phys. Rev. A 65,
022314 (2002).

\bibitem{Kuzmich} Y. O. Dudin, S. D. Jenkins, R. Zhao, D. N.
Matsukevich, A. Kuzmich, and T. A. B. Kennedy, Phys. Rev. Lett.
103, 020505 (2009).

\bibitem{LabeyrieReview} G.Labeyrie, Mod. Phys. Lett. B 22, 73 (2008).

\bibitem{LPLReview} D.V. Kupriyanov, I.M. Sokolov, C.I. Sukenik, and M.D.
Havey, Laser Phys. Lett. 3, 223 (2006).

\bibitem{CommentAMO} Mark D. Havey and Dmitriy V. Kupriyanov,
Phys. Scr. 72, C30 (2005).

\bibitem{KaiserHavey} R. Kaiser and M.D. Havey, Optics and Photonics
News 16, 38 (2005).

\bibitem{Anderson} P. W. Anderson, Phys. Rev. 109, 1492 (1958).

\bibitem{Akkermans1} E. Akkermans, A. Gero, and R. Kaiser, Phys. Rev. Lett. 101, 103602
(2008).

\bibitem{KupriyanovLightStrong} I.M. Sokolov, M.D. Kupriyanova,
D.V. Kupriyanov, and M.D. Havey, Phys. Rev. A 79, 053405 (2009); A. S. Sheremet, A. D. Manukhova, N. V. Larionov, and D. V. Kupriyanov Phys Rev A 86, 043414 (2012)

\bibitem{Fofanov} Ya. A. Fofanov, A.S. Kuraptsev, I.M. Sokolov, and M.D. Havey, Phys. Rev. A 84, 53811 (2011).

\bibitem{JETP} I.M. Sokolov, D.V. Kupriyanov, and M.D. Havey, JETP 112, 246 (2011).

\bibitem{Wiersma1} D.S. Wiersma, P. Bartolini, Ad Lagendijk, and R. Righini,
Nature 390, 671 (1997).

\bibitem{Chabanov1} A.A. Chabanov, M. Stoytchev, and A.Z. Genack,Nature 404, 850 (2000).

\bibitem{Maret1} M. Storzer, P. Gross, C.M. Aegerter, and G. Maret,
Phys. Rev. Lett. 96, 063904 (2006).

\bibitem{Maret} C.M. Aegerter and G. Maret, \emph{Coherent backscattering
and Anderson localization of light}, in Progress in Optics 52, 1
(2009).

\bibitem{Scully} Marlin O. Scully, Phys. Rev. Lett. 102, 143601 (2009).

\bibitem{Svidzinsky} A.A. Svidzinsky, J. Chang, and M.O. Scully, Phys. rev. Lett. 100, 160504 (2008).

\bibitem{KaiserCoop} L. Froufe-Pérez, W. Guerin, R. Carminati, and R. Kaiser
Phys. Rev. Lett, 102, 173903 (2009).

\bibitem{KaiserCoop2} T. Bienaime, R. Bachelard, N. Piovella and R. Kaiser,
Fortschr. Phys. doi: 10.1002/prop.201200089 (2012).

\bibitem{KaiserCoop3} T. Bienaime, M. Petruzzo,D. Bigerni, N. Piovella and R. Kaiser,
J. Mod. Opt. 58, 1942 (2011).

\bibitem{KaiserCoop4} S. Bux, E. Lucioni, H. Bender, T. Bienaime, K. Lauber, C. Stehle, C. Zimmermann, S. Slama, Ph.W.Courteille, N. Piovella and R. Kaiser, J. Mod. Opt. 57, 1841 (2010).

\bibitem{KaiserCoop5} T. Bienaime, S. Bux, E. Lucioni, Ph.W. Courteille, N. Piovella, R. Kaiser
Phys. Rev. Lett. 104, 183602 (2010).

\bibitem{KaiserCoop6} Ph.W. Courteille, S. Bux, E. Lucioni, K. Lauber, T. Bienaime, R. Kaiser, N. Piovella
Eur. Phys. J. D 58, 69 (2010).

\bibitem{Cao} Hui Cao, \emph{Lasing in Disordered Media}, in Progress in Optics 45,
(2003).

\bibitem{Wiersma} D.S. Wiersma, Nature Phys. 4, 359 (2008).

\bibitem{Conti} C. Conti and A. Fratalocchi, Nature Phys. 4, 794 (2008).

\bibitem{KaiserRandom1} L. Froufe-P\`{e}rez, W. Guerin, R. Carminati and R. Kaiser Phys. Rev.
Lett.102, 173903 (2009).

\bibitem{KaiserRandom2} W. Guerin, N. Mercadier, D. Brivio and R. Kaiser,
Optics Exp. 17, 14 (2009).

\bibitem{KaiserRandom3} W. Guerin, N. Mercadier, F. Michaud, D. Brivio, L. S. Froufe-Pérez, R. Carminati, V. Eremeev, A. Goetschy, S. E. Skipetrov, R. Kaiser, J. Opt. A 12, 024002 (2010).

\bibitem{Nielsen} M.A. Nielsen and I.L. Chuang, \emph{Quantum Computation and Quantum Information} (Cambridge University Press, 517,2000).

\bibitem{Briegel} H.-J. Briegel, W. Dur, J.I. Cirac and P. Zoller, Phys. Rev. Lett. 81, 5932 (1998).

\bibitem{Duan} L.-M. Duan, M.D. Lukin, J.I. Cirac and P. Zoller, Nature 414, 413 (2001).

\bibitem{KaiserSubradiant} T. Bienaime, N. Piovella and R. Kaiser, Phys. Rev. Lett. 108, 123602 (2012).

\bibitem{Datsyuk1} V.M. Datsyuk, I.M. Sokolov,D.V. Kupriyanov, and M.D. Havey, Phys. Rev. A 77, 033823 (2008).

\bibitem{Datsyuk2} V.M. Datsyuk, I.M. Sokolv, D.V. Kupriyanov, and M.D. Havey, Phys. Rev. A 74, 043812 (2006).

\bibitem{qm1} I.M. Sokolov, D.V. Kupriyanov, R.G. Olave, and M.D. Havey, J. Modern Optics 57, 1833 (2010).

\bibitem{qm2} L.V. Gerasimov, I.M. Sokolov, D.V. Kupriyanov, R.G. Olave, and M.D. Havey, JOSA B 28, 1459 (2011).

\bibitem{qm3} L.V. Gerasimov, I.M. Sokolov, D.V. Kupriyanov, and M.D. Havey,
J.Phys. B: At. Mol. Opt. Phys. 45 124012 (2012).

\bibitem{Balik23} S. Balik, A.L. Win, M.D. Havey, I.M. Sokolov, and D.V. Kupriyanov, arXiv:0909.1133v3 [quant-ph].


\bibitem{parametric1} S. Balik, A.L. Win, and M.D. Havey, Phys.
Rev. A 80, 023404 (2009).

\bibitem{QUESTFormulas}For an ellipsoidal Gaussian atom
distribution of sizes $r_{o}$ and $z_{o}$ and peak density
$n_{o}$, $n(r) = n_{o}e^{-r^{2}/2r_{o}^{2}-z^{2}/2z_{o}^{2}}$. The
total number of atoms is $N = (2\pi )^{3/2}n_{o}r_{o}^{2}z_{o}$
and the peak transverse (longitudinal) optical depth is
$b_t=\sqrt{2\pi}n_o\sigma_or_o$ ($b_l=\sqrt{2\pi}n_o\sigma_oz_o$).
$\sigma_o$ is the weak field resonance light scattering cross
section.  The peak total cross-section is given by
$\frac{2F^{\prime} + 1}{2F+1}\frac{\lambda^{2}}{2\pi}$.

\bibitem{Scaling} I.M. Sokolov, A.S. Kuraptsev , D.V. Kupriyanov, M.D. Havey and S. Balik
J. Modern Optics, DOI:10.1080/09500340.2012.733431 (2012).


\end{thebibliography}
\end{document}